\documentclass{aastex631}
\usepackage{physics}
\usepackage{CJKutf8}
\usepackage{booktabs}
\pdfoutput=1

\begin{document}
\begin{CJK*}{UTF8}{gbsn}

\title{Polarization degree of magnetic field structure changes caused by random magnetic field in Gamma-ray Burst}

\correspondingauthor{Hong-Bang Liu}
\email{E-mail: liuhb@gxu.edu.cn}

\author[0009-0002-3780-892X]{Jiang-Chuan Tuo (庹江川)}
\affiliation{School of Physical Science and Technology, Guangxi University, Nanning 530004, China}

\author{Hong-Bang Liu}
\affiliation{School of Physical Science and Technology, Guangxi University, Nanning 530004, China}

\author{Qian-Nan Mai}
\affiliation{School of Physical Science and Technology, Guangxi University, Nanning 530004, China}

\author{Qian Zhong}
\affiliation{School of Physical Science and Technology, Guangxi University, Nanning 530004, China}

\author{Zu-Ke Feng}
\affiliation{School of Physical Science and Technology, Guangxi University, Nanning 530004, China}

\author{Kang-Fa Cheng}
\affiliation{School of Mathematics and Physics, Guangxi Minzu University, Nanning 530006, China}

\author[0000-0002-7044-733X]{En-Wei Liang}
\affiliation{School of Physical Science and Technology, Guangxi University, Nanning 530004, China}

\begin{abstract}
In a Poynting-flux-dominated (PFD) jet that exhibits an ordered magnetic field, a transition towards turbulence and magnetic disorder follows after magnetic reconnection and energy dissipation during the prompt emission phase. In this process, the configuration of the magnetic field evolves with time, rendering it impossible to entirely categorize the magnetic field as ordered. Therefore, we assumed a crude model that incorporates a random magnetic field and an ordered magnetic field, and takes into account the proportionality of the random magnetic field strength to the ordered magnetic field, in order to compute the polarization degree (PD) curve for an individual pulse. It has been discovered that the random magnetic field has a significant impact on the PD results of the low-energy X-ray. In an ordered magnetic field, the X-ray segment maintains a significant PD compared to those in the hundreds of keV and MeV ranges even after electron injection ceases, this making PD easier to detect by polarimetry. However, when the random magnetic field is introduced, the low-energy and high-energy PDs exhibit a similar trend, with the X-ray PD being lower than that of the high-energy segment. Of course, this is related to the rate of disorder in the magnetic field. Additionally, there is two rotation of the polarization angles (PAs) that were not present previously, and the rotation of the PA in the high-energy segment occurs slightly earlier. These results are unrelated to the structure of the ordered magnetic field.
\end{abstract}


\section{Introduction} \label{sec:intro}

Gamma-ray Bursts (GRBs) are the most strong explosive phenomenon at cosmological distances. At present, significant progress has been made in the study of GRBs, but there are still many mysteries to be solved. For example, the central engine of GRBs could be a magnetar or a black hole; the exact nature is currently unknown. And these two different central engines will lead to different magnetic field configurations (MFCs) \citep{Spruit2001}, which will affect the evolution of polarization degree (PD). The jet could be either matter-flux-dominated or Poynting-flux-dominated (PFD), it remains a subject of investigation. Early standard fireball model \citep{Paczynski1986,Piran1999,Granot1999,Rees1992} tended to believe that the jet was dominated by matter flow. In this model, strong magnetic fields are generated by internal shocks through Richtmyer-Meshkov instability \citep{Richtmyer1960, Meshkov1972} and Weibel instability \citep{Weibel1959}, and then GRB is generated through synchrotron radiation. However, due to the small coherence scale of these magnetic fields, the PD generated is low, which does not match the data from polarization detectors such as POLAR \citep{ZhangSN2019, Kole2020}, GAP \citep{GAP2012}, and AstroSat \citep{AstroSat2019}. Therefore, it is reasonable to speculate that the jet may be slightly magnetized or highly magnetized \citep{Thompson1994,Drenkhahn2002,Beniamini2017,Zhangbing2011}. \cite{Gao2015} proposed a more generalized hybrid jet model that incorporates both entropy and magnetization. This model is capable of explaining the coexistence of thermal and non-thermal components observed in some GRBs.

Typically, the energy spectrum of GRB is fitted with the Band spectrum \citep{Band1993}, and the typical low-energy spectrum index and high-energy spectrum index are $\alpha=-1$ and $\beta=-2.2$, respectively. The radiation mechanism of GRB is still unknown. 
\cite{Uhmandzhang2014} proposed a synchrotron radiation model of electrons in a decaying magnetic field to generate the Band spectrum. \cite{Geng2018} and \cite{WangK2021} studied the energy spectrum in the case of magnetic field decay, and \cite{Zhang2016} conducted a time-resolved spectral analysis of GRB 130606B, which showed that it could not be distinguished from the Band spectrum within the detection energy range of 8 keV to 40 MeV of \emph{Fermi}/GBM \citep{Meegan2009}.

The polarization generated by different GRB models will show different time and energy dependencies, leading to different PDs and polarization angles (PAs). Therefore, by observing the time-resolved of PDs and statistics of its distribution, we can effectively distinguish different GRB models. In the case of an ordered magnetic field, synchrotron radiation exhibits high PD in different energy bands \citep{Toma2009,Gill2020}. However, in the case of a random magnetic field, if $1/\Gamma$ cone is within the geometric cone of the jet, the PD drops to 0, where $\Gamma$ represent bulk Lorentz factor. Polarization can only be observed when the condition $\theta_\mathrm{V}-1/\Gamma<\theta_\mathrm{J}<\theta_\mathrm{V}+1/\Gamma$ is met due to symmetry breaking \citep{Ghisellini1999,Granot2003,Toma2009,Gill2020}, where $\theta_\mathrm{J}$ and $\theta_\mathrm{V}$ represent the jet opening angle and the angle between the line of sight (LOS) and the jet axis, respectively. For the Compton drag model, the situation is similar to that of the random magnetic field \citep{Lazzati2004}. On the other hand, the photosphere model proposes that as the energy band increases, the PD decreases accordingly, and the polarization phenomenon disappears in energy band higher than $E_\mathrm{P}$ \citep{Lundman2014,Lundman2018}.

POLAR had successfully captured the polarization properties of 14 GRBs with a detection energy range of 50 keV to 500 keV. These results of GRBs' polarization characteristics indicate that the PD of GRB is relatively low \citep{ZhangSN2019, Kole2020}. GRB 170114A is particularly noteworthy, as a $90^\circ$ flip in the PA was observed in this GRB \citep{ZhangSN2019}. A further analysis by \cite{Burgess2019} suggests that GRB 170114A may have experienced two $90^\circ$ flips in PA. However, the result has a large error. Therefore, a polarization detector with greater accuracy was necessary. As a follow-up task of POLAR, POLAR-2 \citep{Feng2023, Angelis2022} will further expand the measurement energy range and improve measurement accuracy. The plan is to deploy it on the Chinese Space Station in 2026. POLAR-2 will conduct time-resolved observations of PDs and PAs for bright GRBs, enabling differentiation among various GRB models.

In this work, we aim to investigate the evolution of MFC in individual pulse of GRB prompt emission. Specifically, we focus on the evolution of the ratio between ordered and disordered magnetic fields over time, and how this affects the PD and PA at different energy bands. The paper is arranged as follows. In Section \ref{sec:model}, we present our model. In Section \ref{sec:result}, we present the calculated polarization properties. Finally, we summarize and discuss our results in Section \ref{sec:Conclusions}.

\section{model}\label{sec:model}
To effectively describe the magnetic field, three right-handed coordinate systems are established to obtain the PD and PA \citep{Sari1999,Toma2009,Lan2018,Lan2019}: The global coordinate system $\hat{X}\hat{Y}\hat{k}$, the direction of $\hat{k}$ aligns with the LOS direction，$\hat{Y} = \cfrac{\hat{k} \times \hat{J}}{|\hat{k} \times \hat{J}|}$ ($\hat{J}$ is the direction of the jet axis), and the direction of $\hat{X}$, which is given by $\hat{X} = \hat{Y}\times\hat{k}$, aligns with the projection of $\hat{J}$ onto the sky plane. Hereafter, the superscript hat will be employed to represent unit vectors. The local fluid frame coordinate system $\hat{x}\hat{y}\hat{\beta}$, the direction of $\hat{\beta}$ corresponds to the direction of the local fluid velocity, which can be represented in the $\hat{X}\hat{Y}\hat{k}$ coordinate system as $(\sin\theta\cos\phi, \sin\theta\sin\phi, \cos\theta)$, where $\theta$ is the inclination angle between the direction of $\hat{k}$-axis and the direction of the $\hat{\beta}$-axis and $\phi$ is the azimuthal angle with the $\hat{X}$-axis, $\hat{y}=\cfrac{\hat{\beta} \times \hat{k}}{|\hat{\beta} \times \hat{k}|}$, and $\hat{x}=\hat{y}\times\hat{\beta}$. The local comoving frame coordinate system $\hat{1} \hat{2}\hat{k}^\prime$, $\hat{k}'=\cfrac{1}{\Gamma(1-\beta\cdot\hat{k})}\left[\hat{k}+\beta\left(\cfrac{\Gamma^{2}}{\Gamma+1}\beta\cdot\hat{k}-\Gamma\right)\right]$ represents the direction of $\hat{k}$ in the comoving frame, where $\beta = \cfrac{\sqrt{\Gamma^2-1}}{\Gamma}$ with the bulk Lorentz factor of the shell, $\hat{1}=\hat{y}$, and $\hat{2}=\hat{k'}\times\hat{1}$.

\subsection{The Model of The Magnetic Field}
The radiation mechanism of GRB is not understood, synchrotron radiation occurring at large radii from central engine is one possible model, such as the Internal-Collision-Induced Magnetic Reconnection and Turbulence (ICMART) model \citep{Zhangbing2011}. To achieve a band-function-like form in the computed fast-cooling synchrotron radiation, this work employs the decay law for magnetic field strength as proposed by \citep{Uhmandzhang2014}, which is given by
\begin{equation}
    B^\prime=B_\mathrm{0}^\prime(\frac{R}{R_\mathrm{0}})^{-b},
    \label{eq:MF}
\end{equation}
where $B'$ is the magnetic field strength in comoving frame, b is the magnetic field strength decaying index, and fixes the values of $R_0 = 10^{15}$ cm and $B'_0= 30$ G. In this paper, the primed quantities are expressed in the comoving frame. In general, the occurrence of magnetic reconnection gives rise to magnetic dissipation, ultimately resulting in a value of $ b $ greater than 1. Nevertheless, the specific value of $ b $ does not have a significant impact on the fluctuations observed in PD and PA. Consequently, for the purposes of our analysis, we have elected to set $ b $ equal to 1 as a constant value.

In a smaller region，random magnetic field can be considered as having a fixed direction. And according to \cite{Lan2019} the mixed magnetic field, within the $\hat{x}\hat{y}\hat{\beta}$ coordinate system, the total magnetic field can be expressed as
\begin{align}
    &\hat{B}_\mathrm{total}=\cfrac{|B_\mathrm{ord}|}{|B_\mathrm{total}|}\hat{B}_\mathrm{ord}+\cfrac{|B_\mathrm{rand}|}{|B_\mathrm{total}|}|\hat{B}_\mathrm{rand},\\
    &|B_\mathrm{total}|=\eta|B_\mathrm{rand}|,\\
    &\eta=\sqrt{1+\xi^2+2\xi_\mathrm{B} \hat{B}_\mathrm{ord}\cdot\hat{B}_\mathrm{rand}},\\
    &\xi_\mathrm{B} = |B_\mathrm{ord}|/|B_\mathrm{rand}|,
\end{align}
where $\xi_\mathrm{B}$ is the ratio of ordered to random magnetic field strength. We make the assumption that the orientation of the random magnetic field component of the total magnetic field lies within the $(\hat{x},\hat{y})$ plane, and follows an isotropic distribution. Finally, the mixed magnetic field in a smaller region can be represented as $\hat{B}_\mathrm{total}=(B_\mathrm{total,x},B_\mathrm{total,y},0)$, and
\begin{equation}
\begin{aligned}
&B_\mathrm{total,x} = (\xi_\mathrm{B} B_\mathrm{ord,x}+\cos\phi_\mathrm{r})/\eta,  \\
&B_\mathrm{total,y} = (\xi_\mathrm{B} B_\mathrm{ord,y}+\sin\phi_\mathrm{r})/\eta.
\end{aligned}
\end{equation}
Subsequently, we can determine the average direction of the mixed magnetic field in the local fluid element within the comoving frame, i.e., $\overline{\sin\theta_\mathrm{B}'}$，$\overline{\sin{\theta_\mathrm{B}'}\sin{2\phi_\mathrm{B}'}}$，$\overline{\sin{\theta_\mathrm{B}'}\cos{2\phi_\mathrm{B}'}}$.

In the ICMART model, it is postulated that jet is highly magnetized. The outflow is visualized as a series of shells, in which the magnetic field exhibits a highly ordered state. Due to elastic collisions between these shells, magnetic reconnection and turbulence phenomena occur, which can disrupt the originally ordered magnetic field. Therefore, we believe that there is a process of transition from an ordered state to a disordered state in the magnetic field. To facilitate a parametric estimation of the temporal evolution of polarization, we postulate a simplified assumption: the ratio of ordered to random magnetic field strength can be described by
\begin{equation}
    \xi_\mathrm{B}(R)=\xi_\mathrm{B,0}(\frac{R}{R_\mathrm{0}})^{-\kappa},
    \label{eq:E}
\end{equation}
where $\kappa$ is a power-law decay index. We can utilize the parameter $\kappa$ to regulate the rate of decay, and specifically, when $\kappa$ equals zero, it corresponds to a constant magnetic field strength ratio. In reality, the transformation of $\xi_\mathrm{B}$ does not necessarily follow a simple power law form. A more reasonable evolution of magnetic fields over time needs to be derived from numerical simulations. And preliminary investigations have been conducted by \cite{Deng2015,Deng2016}.

According to \cite{Lan2018,Lan2019}, the equation for the pure toroidal MFC is given by
\begin{equation}
\begin{aligned}
    &\hat{B}_\mathrm{T}' = J_\mathrm{T,x}/A_\mathrm{T}\hat{x}+J_\mathrm{T,y}/A_\mathrm{T}\hat{y},\\
    &J_\mathrm{T,x} = -\sin{\theta_\mathrm{V}}\sin{\phi},\\
    &J_\mathrm{T,y} = -\sin{\theta_\mathrm{V}}\cos{\theta}\cos{\phi}+\cos{\theta_\mathrm{V}}\sin{\theta},\\
    &A_\mathrm{T} = \sqrt{J_\mathrm{T,x}^2+J_\mathrm{T,y}^2},
\end{aligned}
\end{equation}
and the formula we use for the purely aligned MFC is
\begin{equation}
\begin{aligned}
    &\hat{B}_\mathrm{A}'=J_\mathrm{A,x}/A_\mathrm{A}\hat{x}+J_\mathrm{A,y}/A_\mathrm{A}\hat{y},\\
    &J_\mathrm{A,x}=-\sin{\phi}\cos{\theta_\mathrm{V}}\sin{\delta_\mathrm{a}}-\cos{\phi}\cos{\delta_\mathrm{a}},\\
    &J_\mathrm{A,y}=\cos{\theta}\sin{\phi}\cos{\delta_\mathrm{a}}-\sin{\theta}\sin{\theta_\mathrm{V}}\sin{\delta_\mathrm{a}}-\cos{\theta}\cos{\phi}\cos{\theta_\mathrm{V}}\sin{\delta_\mathrm{a}},\\
    &A_\mathrm{A} = \sqrt{J_\mathrm{A,x}^2+J_\mathrm{A,y}^2},
\end{aligned}
\end{equation}
where $\delta_\mathrm{a}$ is the angle between the direction of magnetic field and $\hat{X}$-axis. 

\subsection{Evolution of the Electron Spectrum}

In our model, we assume magnetic reconnection occurs at $R_{\mathrm{on}}$ and ends at $R_{\mathrm{off}}$. The energy released by magnetic reconnection directly accelerates electrons, while turbulence also stochastically accelerates electrons, ultimately resulting in a powerlaw distribution electron, and the injection electron spectrum can be written as 
follow
\begin{equation}\label{eq:JS}
    Q'(\gamma_\mathrm{e}')=C\gamma_\mathrm{e}'^{-p}\quad (\gamma_{\mathrm{min}}'<\gamma_\mathrm{e}'<\gamma_{\mathrm{max}}'),
\end{equation}
where $\gamma_{\mathrm{min}}'$ and $\gamma_{\mathrm{max}}'$ are the min Lorentz factor and the maximum Lorentz factor, respectively.
$\gamma_{\mathrm{max}}'$ is dependent on the magnetic field, and the relationship of $\gamma_{\mathrm{max}}'$ with the magnetic field is \citep{Dai1998}
\begin{equation}
    \gamma_{\mathrm{max}}'=(\cfrac{6\pi e}{\sigma_\mathrm{T} B'})^{1/2},
\end{equation}
where e is the electron charge, and $\sigma_\mathrm{T}$ is the Thomson scattering cross section. The electron radiate synchrotron photons, and the arrival of the first photon at the observer is considered of the initial moment of the prompt emission.

In the magnetic field environment described by equation \ref{eq:MF}, the electron spectrum injected per second is known. The electron spectrum can be obtained by solving the continuity equation of the electrons in energy space
\begin{equation}
\frac{\partial}{\partial t'}\left(\frac{\mathrm{d}N_{\mathrm{e}}^{\prime}}{\mathrm{d}\gamma_{\mathrm{e}}^{\prime}}\right)+\frac{\partial}{\partial\gamma_{\mathrm{e}}^{\prime}}\left[\dot{\gamma}_{\mathrm{e,tot}}^{\prime}\left(\frac{\mathrm{d}N_{\mathrm{e}}^{\prime}}{\mathrm{d}\gamma_{\mathrm{e}}^{\prime}}\right)\right]=Q'(\gamma_{\mathrm{e}}^{\prime}),
\end{equation}
where $\frac{\mathrm{d}N_{\mathrm{e}}^{\prime}}{\mathrm{d}\gamma_{\mathrm{e}}^{\prime}}$ is the instantaneous electron spectrum of the system at the epoch $t^{\prime}$, $\dot{\gamma}_{\mathrm{e,tot}}^{\prime}$ is the total cooling rate of the electrons, and 
 $Q'(\gamma_{\mathrm{e}}^{\prime})$ is the source function as defined in equation \ref{eq:JS}. The calculation method is detailed in references \citep{ChangandCooper1970, Chiaberge1999, Cheng2020,Geng2018}.

In this paper, we have not considered the influence of inverse Compton cooling on the electron spectrum. This is because inverse Compton cooling is dependent on the magnetic field, and this study primarily focuses on the behavior of the jet under the PFD jet in the ICMART model. Therefore, we have neglected the effects of inverse Compton cooling in this scenario.

\subsection{Calculation Method of Polarization}

In the $\hat{1} \hat{2}\hat{k}^\prime$ coordinate system, the direction of the magnetic field $\hat{B}'$ can be represented as $(\sin\theta_\mathrm{B}'\cos\phi_\mathrm{B}', \sin\theta_\mathrm{B}'\sin\phi_\mathrm{B}', \cos\theta_\mathrm{B}')$, where $\theta_\mathrm{B}'$ is the pitch angle between the direction of $\hat{B'}$ and the direction of the $k'$-axis and $\phi_\mathrm{B}'$ is the azimuthal angle with the $\hat{1}$-axis, the synchrotron emission power emitted per unit frequency by an electron with Lorentz factor $\gamma_\mathrm{e}$ in a magnetic field of magnitude $B^{\prime}$ is \citep{RybickiandLightman1979}
\begin{equation}
p'(\nu')=\frac{\sqrt3e^3 B'\sin\theta_\mathrm{B}'}{m_\mathrm{e} c^2}F\Bigg(\frac{\nu'}{\nu_\mathrm{ch}'}\Bigg),
\end{equation}
where $m_\mathrm{e}$ is the rest mass of the electron, c is the speed of light, $F(x)= x\int_{x}^{+\infty}K_{5/3}(k)dk$, where $x=\cfrac{\nu'}{\nu'_\mathrm{ch}}$, is the dimensionless spectrum of synchrotron radiation with the modified Bessel function of $\cfrac{5}{3}$ order, and $\nu_\mathrm{ch}^{\prime} = \cfrac{3}{4\pi}\cfrac{eB'\sin\theta_\mathrm{B}^{\prime}}{m_\mathrm{e} c}\gamma_\mathrm{e}'^{2}$ is the characteristic frequency. 

In a local point, the Stokes parameters can be expressed as $F^\prime_\mathrm{p}=A_\mathrm{0} B'\sin\theta_\mathrm{B}^\prime \int_{\gamma'_\mathrm{e,min}}^{\gamma'_\mathrm{e,max}}F(x)N'(\gamma_\mathrm{e}')d\gamma_\mathrm{e}'$, $Q^\prime_\mathrm{p}=F'_\mathrm{p}\Pi_\mathrm{0}\cos{2\chi^\prime_\mathrm{p}}$, $U^\prime_\mathrm{p}=F'_\mathrm{p}\Pi_\mathrm{0}\sin{2\chi^\prime_\mathrm{p}}$, where $A_\mathrm{0}$ is a constant at a fixed position in the jet, and $B'$ is not regarded as a constant value due to taking into account the random magnetic field. The PA $\chi'_\mathrm{p}=\phi_\mathrm{B}'-\cfrac{\pi}{2}$, so that 
\begin{equation}
    Q^\prime_\mathrm{p}=-A_\mathrm{0}B'\sin\theta_\mathrm{B}^\prime \Pi_\mathrm{0}\cos{2\phi_\mathrm{B}'} \int_{\gamma'_\mathrm{e,min}}^{\gamma'_\mathrm{e,max}}F(x)N'(\gamma_\mathrm{e}')d\gamma_\mathrm{e}', 
\end{equation}
\begin{equation}
    U^\prime_\mathrm{p}=-A_\mathrm{0}B'\sin\theta_\mathrm{B}^\prime \Pi_\mathrm{0}\sin{2\phi_\mathrm{B}'} \int_{\gamma'_\mathrm{e,min}}^{\gamma'_\mathrm{e,max}}F(x)N'(\gamma_\mathrm{e}')d\gamma_\mathrm{e}',
\end{equation}
and $\int_{\gamma'_\mathrm{e,min}}^{\gamma'_\mathrm{e,max}}F(x)N'(\gamma_\mathrm{e}')d\gamma_\mathrm{e}'$ does not depend on the direction of the magnetic field. The local PD and PA in comoving frame can be expressed as 
\begin{equation}
\begin{aligned}
    \Pi_\mathrm{p}' =\cfrac{\sqrt{\langle Q^\prime_\mathrm{p}\rangle^2+\langle U^\prime_\mathrm{p}\rangle^2}}{\langle F^\prime_\mathrm{p}\rangle}, \chi_\mathrm{p}'=\cfrac{1}{2}\arctan{\cfrac{\langle U^\prime_\mathrm{p}\rangle}{\langle Q^\prime_\mathrm{p}\rangle}}.
\end{aligned}
\end{equation}

The local PD in $\hat{X}\hat{Y}\hat{k}$ coordinate system can be expressed as $\Pi_\mathrm{p}=\Pi_\mathrm{p}'\propto\Pi_\mathrm{0}$, where $\Pi_\mathrm{0}$ represents the intrinsic PD of the electron ensemble, which, for non-powerlaw electron spectrum, is given by
\begin{equation}
    \Pi_\mathrm{0}=\cfrac{\int_{\gamma_\mathrm{e,min}'}^{\gamma_\mathrm{e,max}'}G(x)N'(\gamma_\mathrm{e}')d\gamma_\mathrm{e}'}{\int_{\gamma_\mathrm{e,min}'}^{\gamma_\mathrm{e,max}'}F(x)N'(\gamma_\mathrm{e}')d\gamma_\mathrm{e}'}.
\end{equation}
where $G(x)=xk_{2/3}(x)$ with the modified Bessel function of $\cfrac{2}{3}$ order.

According to \cite{Lan2019}, the local PA in $\hat{X}\hat{Y}\hat{k}$ coordinate system can be expressed as $\chi_\mathrm{p} = \phi + \chi_\mathrm{p}' + \frac{\pi}{2} + n\pi$, where $n$ is an integer, and $\chi_\mathrm{p}$ takes values within the range $[- \frac{\pi}{2}, \frac{\pi}{2}]$. 

The Stokes parameters observed in the observer frame can be expressed as
\begin{equation}
    F_\mathrm{\nu}=\frac{1+z}{4\pi D_\mathrm{L}^{2}}\int_{0}^{\theta_\mathrm{J}+\theta_\mathrm{V}}d\theta{\mathcal D}^{3}\sin\theta \int_{-\Delta\phi}^{\Delta\phi}p^{\prime}(\nu^{\prime})d\phi,
\end{equation}
\begin{equation}
    Q_\mathrm{\nu}=\frac{1+z}{4\pi D_\mathrm{L}^{2}}\int_{0}^{\theta_\mathrm{J}+\theta_\mathrm{V}}d\theta{\mathcal D}^{3}\sin\theta \int_{-\Delta\phi}^{\Delta\phi}p^{\prime}(\nu^{\prime})\Pi_\mathrm{p}\cos(2\chi_\mathrm{p})d\phi,
\end{equation}
\begin{equation}
    U_\mathrm{\nu}=\frac{1+z}{4\pi D_\mathrm{L}^{2}}\int_{0}^{\theta_\mathrm{J}+\theta_\mathrm{V}}d\theta{\mathcal D}^{3}\sin\theta \int_{-\Delta\phi}^{\Delta\phi}p^{\prime}(\nu^{\prime})\Pi_\mathrm{p}\sin(2\chi_\mathrm{p})d\phi,
\end{equation}
where $P^{\prime}(\nu^{\prime}) = \int N'(\gamma_\mathrm{e}')p^\prime(\nu^\prime)d\gamma_\mathrm{e}'$ is the power of the electron spectrum in $\hat{1} \hat{2}\hat{k}^\prime$ coordinate system, z is cosmological redshift of the source, which is set to 1, $D_\mathrm{L}$ is the luminosity distance of the burst, $\mathcal D=1/(\Gamma(1-\beta \cos{\theta}))$ is the doppler factor, $\nu'=\nu(1+z)/\mathcal D$, and $\Pi_\mathrm{p}$ and $\chi_\mathrm{p}$ are the PD and PA of a local point, respectively. The $\Delta \phi$ can be expressed by \citep{Wu2005}
\begin{equation}
    \Delta\phi=\begin{cases}\pi\Theta(\theta_\mathrm{J}-\theta_\mathrm{V}),&\theta\leqslant\theta_-,\\
    \arccos\left(\cfrac{\cos\theta_\mathrm{J}-\cos\theta_\mathrm{V}\cos\theta}{\sin\theta_\mathrm{V}\sin\theta}\right),&\theta_-<\theta<\theta_+,\\
    0,&\theta\geqslant\theta_+,
    \end{cases}
\end{equation}
where $\theta_-=\abs{\theta_\mathrm{J}-\theta_\mathrm{V}}$, $\theta_+=\theta_\mathrm{J}+\theta_\mathrm{V}$, and $\Theta(x)$ is the Heaviside step function. 

The final PD and PA of the observer can be calculated as follow
\begin{equation}
    \Pi = \cfrac{\sqrt{Q_\mathrm{\nu}^2+U_\mathrm{\nu}^2}}{F_\mathrm{\nu}},
    \label{eq:PD}
\end{equation}
\begin{equation}
    \chi = \cfrac{1}{2}\arctan\cfrac{U_\mathrm{\nu}}{Q_\mathrm{\nu}},
    \label{eq:PA}
\end{equation}
where $\chi$ can be regarded as the angle between the observed total electric vector and $\hat{X}$-axis, and when using Stokes parameters to calculate the $\chi_\mathrm{cal}$, it is important to note that the real $\chi_\mathrm{real}$ is defined as follow
\begin{equation}
    \chi_\mathrm{real} = \begin{cases}
        \chi_\mathrm{cal}, &Q>0,\\
        \chi_\mathrm{cal}+\cfrac{\pi}{2},&Q<0\; \mathrm{and}\; U\geq0,\\
        \chi_\mathrm{cal}-\cfrac{\pi}{2},&Q<0\; \mathrm{and}\; U<0.
    \end{cases}
\end{equation}

Consider an infinitely thin shell. Assuming that the initial magnetic field is contained within the shell, following magnetic reconnection and turbulence, the field gradually becomes entangled but remains confined within the shell. Disregarding the radial structure of the emitting regions, we focus on the shell's properties. Thus, in the $\hat{x}\hat{y}\hat{\beta}$ coordinate system, the magnetic field can be represented as $B=(B_\mathrm{x},B_\mathrm{y},0)$.

Based on the work of \cite{Toma2009}, the mixed magnetic field $\hat{B}_\mathrm{total}=(B_\mathrm{total,x},B_\mathrm{total,y},0)=(\cos{\phi_\mathrm{B}},\sin{\phi_\mathrm{B}},0)$ in the local fluid frame, where $\phi_\mathrm{B}$ is the azimuthal angle with the $\hat{x}$-axis. $\hat{B}_\mathrm{total}$ is transformed from the $\hat{x}\hat{y}\hat{\beta}$ coordinate system to the $\hat{1}\hat{2}\hat{k'}$ coordinate system, resulting in a new representation, $\hat{B}_\mathrm{total}'=(\sin\theta_\mathrm{B}'\cos\phi_\mathrm{B}', \sin\theta_\mathrm{B}'\sin\phi_\mathrm{B}', \cos\theta_\mathrm{B}')$, where $\theta'_\mathrm{B}$ is the pitch angle between the direction of $\hat{k}'$-axis and the direction of the $\hat{B}'_\mathrm{total}$ and $\phi'_\mathrm{B}$ is the azimuthal angle with the $\hat{1}$-axis. We can derive the transformation relationship between $\hat{B}_\mathrm{total}$ and $\hat{B}_\mathrm{total}'$ as follow
\begin{equation}
\begin{aligned}
&\cos\theta_\mathrm{B}^{\prime}=B_\mathrm{total,x}\sin\theta^{\prime}, \\
&\sin\theta_\mathrm{B}^{\prime}=\sqrt{1-(\cos\theta_\mathrm{B}^{\prime})^{2}}, \\
&\sin{\theta_\mathrm{B}'}\sin{2\phi_\mathrm{B}'}= -\cfrac{2B_\mathrm{total,x}B_\mathrm{total,y}\cos{\theta'}}{\sin{\theta_\mathrm{B}'}},\\
&\sin{\theta_\mathrm{B}'}\cos{2\phi_\mathrm{B}'}= \cfrac{2B_\mathrm{total,y}^2}{\sin{\theta_\mathrm{B}'}}-\sin{\theta_\mathrm{B}'}. 
\end{aligned}
\end{equation}

At a burst source time t and angle $ \theta $, where $ \theta $ represents the angle between the region emitting photons and the LOS. Photons emitted at this angle are received at $ t_\mathrm{obs} $ in the observer's frame, and it can be expressed as \citep{Granot1999,Huang2000,Lan2021}

\begin{equation}
    \cfrac{t_\mathrm{obs}}{(1+z)} = (t-t_\mathrm{on})(1-\beta_\mathrm{sh}\cos\theta)+(1-\cos\theta)R_{\mathrm{on}}/c,
\end{equation}
where $\beta_\mathrm{sh}$ is the dimensionless velocity of the shell. In this paper, all the calculations take into account the equal arrival time surface (EATS).

\section{numerical result}\label{sec:result}

For the electron spectrum, we use a model with parameters $R_\mathrm{on} = 10^{14}$ cm, $R_{\mathrm{off}} = 10^{16}$ cm, $\Gamma = 450$ and an constant injection rate $N_\mathrm{inj}'=\int^{\gamma_\mathrm{max}'}_{\gamma_\mathrm{min}'}Q'(\gamma_\mathrm{e}')d\gamma_\mathrm{e}'=5\times10^{45}$ s$^{-1}$.
Additionally, as the PD is related to the MFCs of the shell, we consider both toroidal and aligned MFCs.

In the toroidal MFC, $\sin(2\chi_\mathrm{p})$ exhibits odd symmetry with respect to $\phi$. Consequently, $U_\mathrm{\nu}$ is 0 in the observer frame. This allows the final PD equation \ref{eq:PD} to be simplified to $\Pi=\cfrac{Q_\mathrm{\nu}}{F_\mathrm{\nu}}$. Additionally, for the PA, the sign of PD can be utilized to indicate the direction of PA. When $Q_\mathrm{\nu}$ is greater than 0 (or less than 0), PA is at an angle of $0^\circ$ (or $90^\circ$) and is parallel (or perpendicular) to the $\hat{X}$-axis. For the random magnetic field, $U_\mathrm{\nu}$ is also 0, thus consistent with the representation method for the toroidal MFC.

For the toroidal MFC, PD and PA are strongly associated with the equation $y=(\Gamma \theta_\mathrm{J}) ^ 2$ \citep{Toma2009,Cheng2020,Gill2020}. To investigate the impact of variations in the proportion of random magnetic field on the observed PD and PA, we selected a set of values: $\theta=0.04$, $\Gamma=450$, and $y=324$, which exhibit a pronounced and easily recognizable pattern of PDs and PAs.

\begin{figure}
    \centering
    \includegraphics[scale=0.35]{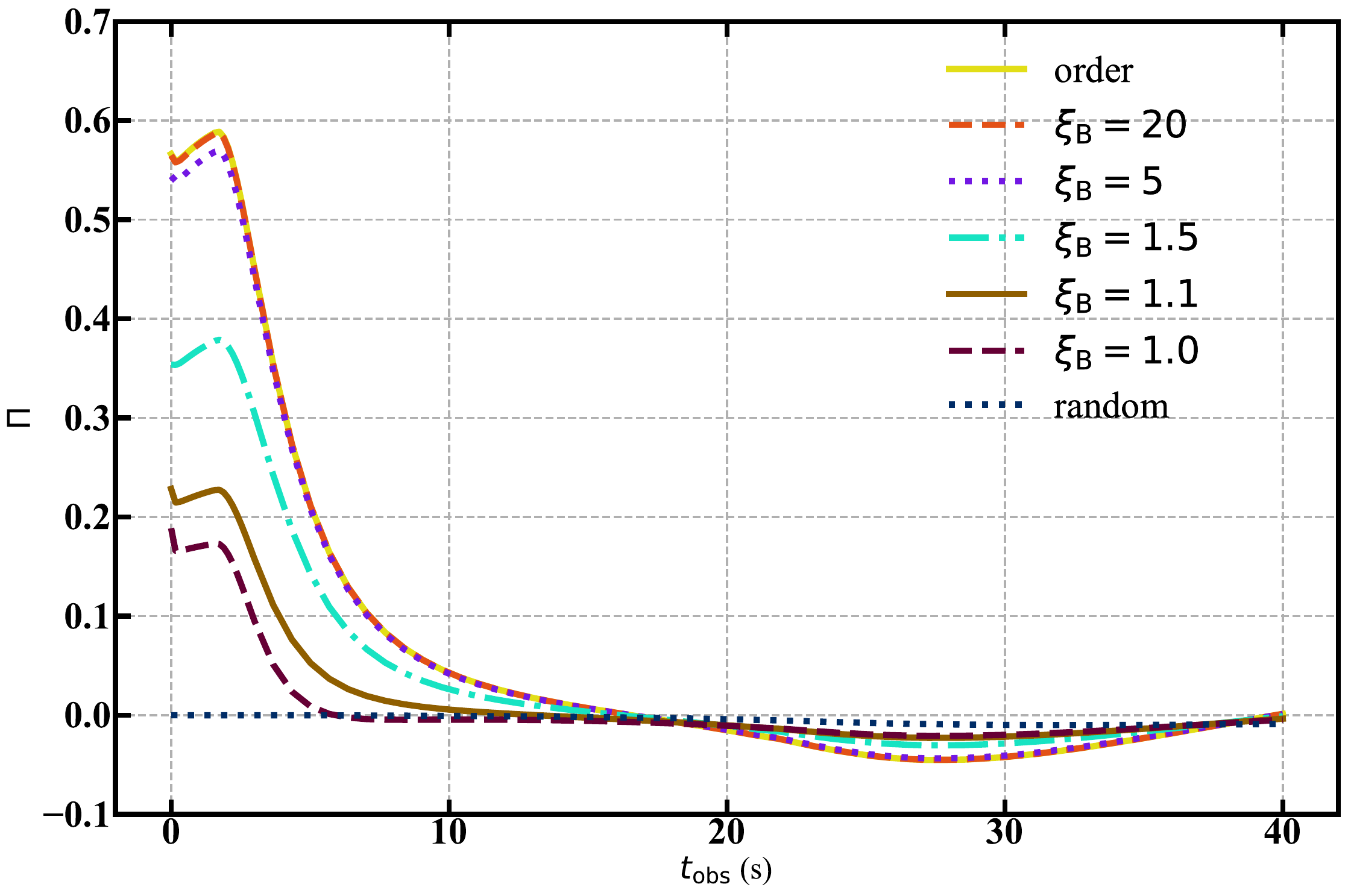}
    \caption{The temporal evolution of polarization under various constant values of $\xi_\mathrm{B}$ while keeping the observing angle $q=0.8$, the ordered implication is $\xi_\mathrm{B}\to \infty$ for the toroidal MFC, the random implication is $\xi_\mathrm{B}= 0$, and the observational energy is taken as $h\nu_\mathrm{obs}=200$ keV.}
    \label{fig:orderandrandom}
\end{figure}

For the ICMART model, the shell is PFD, and the magnetic field is ordered ($\xi_\mathrm{B}\to \infty$) in the early stages of GRBs. The PD and PA are intrinsic properties of the MFC within the shell. As the strength of the random magnetic field increases, the PD decreases. Consequently, the observed PDs and PAs will differ from those observed in an ordered MFC. In general, for order magnetic field, $\xi_\mathrm{B}$ is very larger, and for the random magnetic field, $\xi_\mathrm{B}$ is very smaller. Figure \ref{fig:orderandrandom} depicts the temporal evolution of PD curves under constant values of $\xi_\mathrm{B}$, the ordered implication is $\xi_\mathrm{B}\to \infty$ for the toroidal MFC. The random implication is that $\xi_\mathrm{B} = 0$. The observational energy is taken as $\nu_\mathrm{obs}=200$ keV, and the observing angle $q=0.8$. 

\begin{figure}
    \centering
    \includegraphics[scale=0.35]{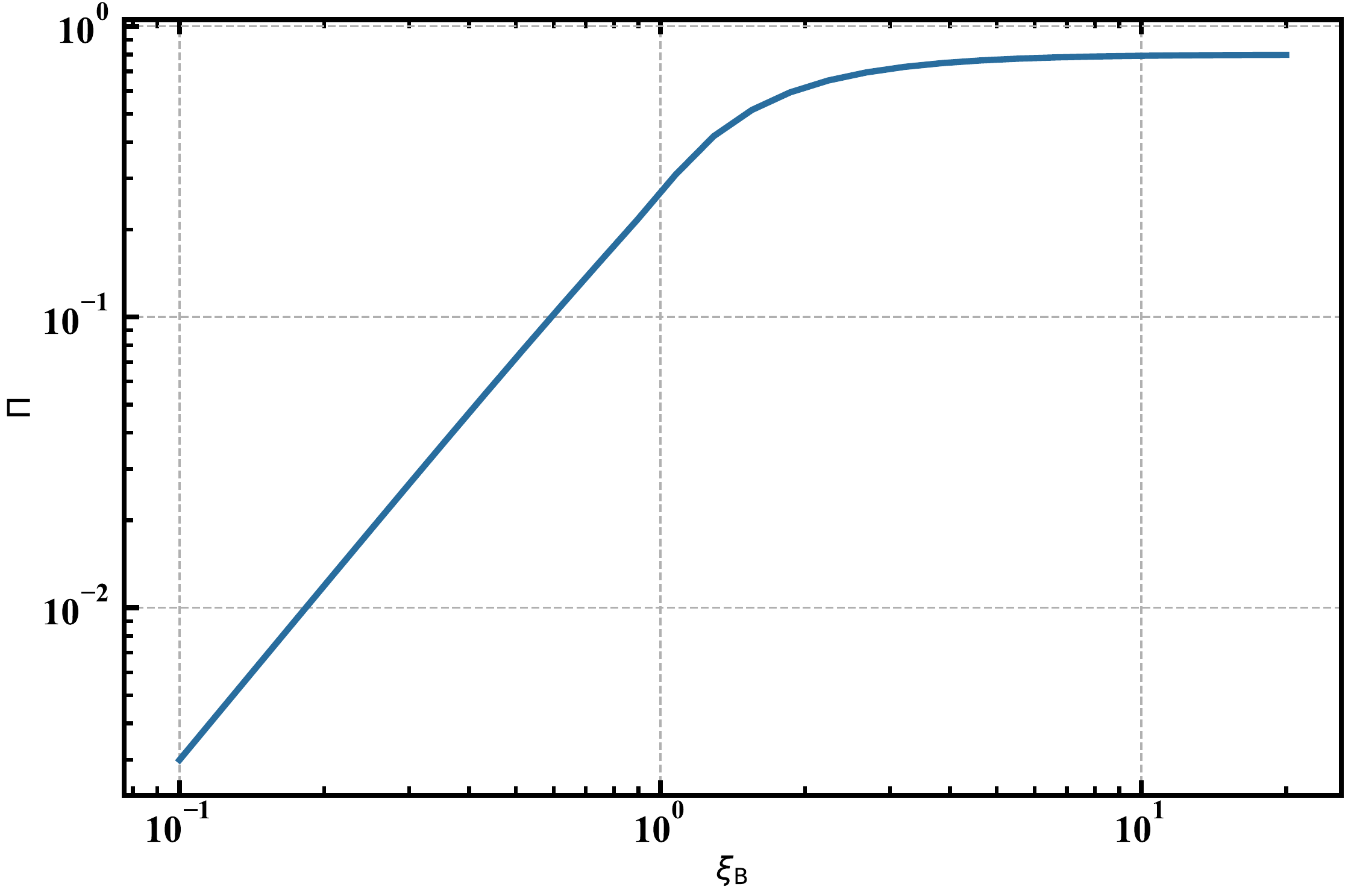}
    \caption{In a given local point, consider the evolution of the PD with respect to $\xi_\mathrm{B}$. Furthermore, assume that the angle between the normal of that local point and the LOS is $0^\circ$, and let the PD be $\Pi=0.8$ in the magnetic configuration under consideration.}
    \label{fig:variousepsilon}
\end{figure}

\begin{figure}[htbp]
	\centering
	\begin{minipage}{0.7\linewidth}
		\centering
		\includegraphics[width=0.9\linewidth]{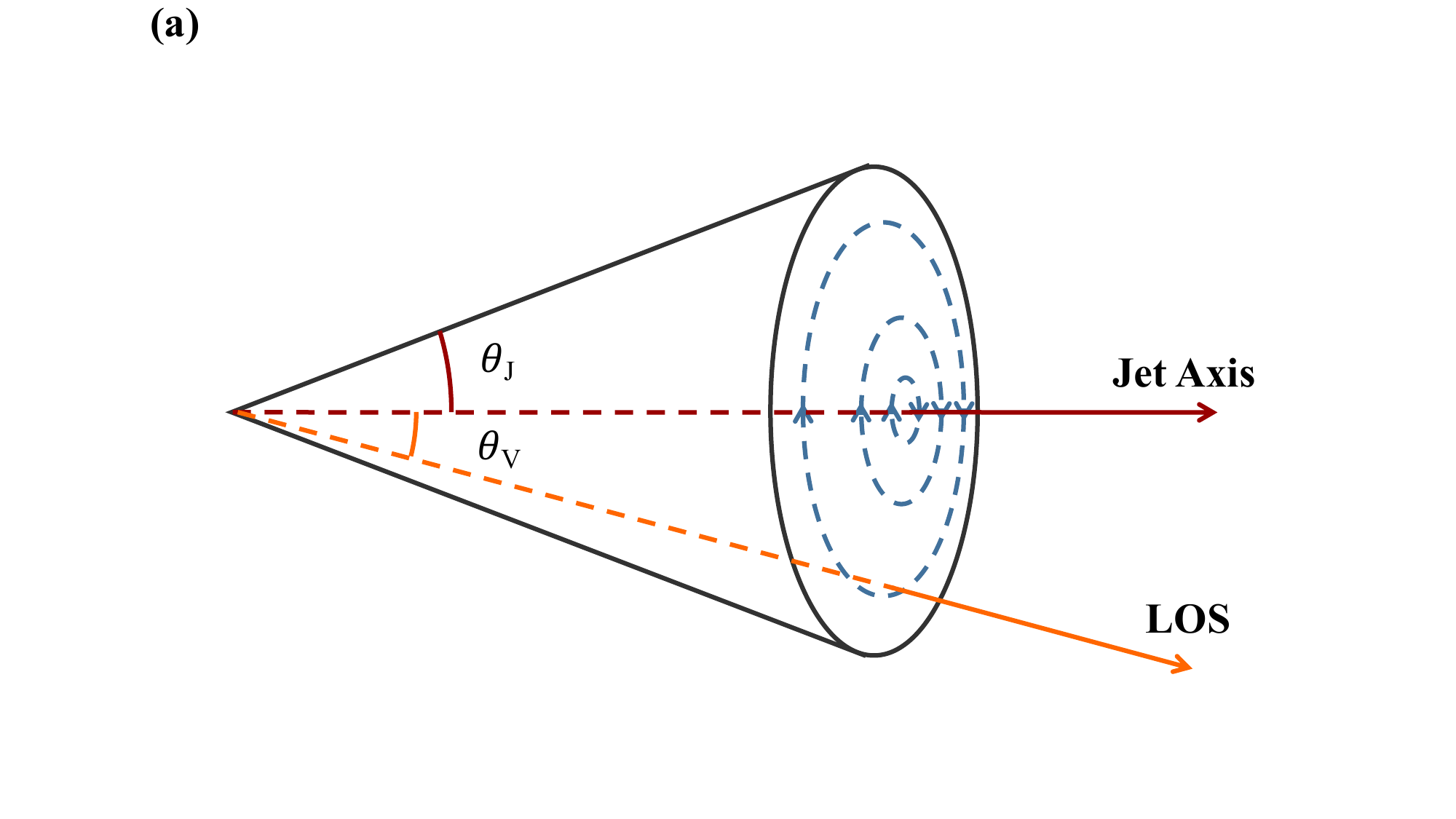}
	\end{minipage}
	\centering
	\begin{minipage}{0.7\linewidth}
		\centering
		\includegraphics[width=0.9\linewidth]{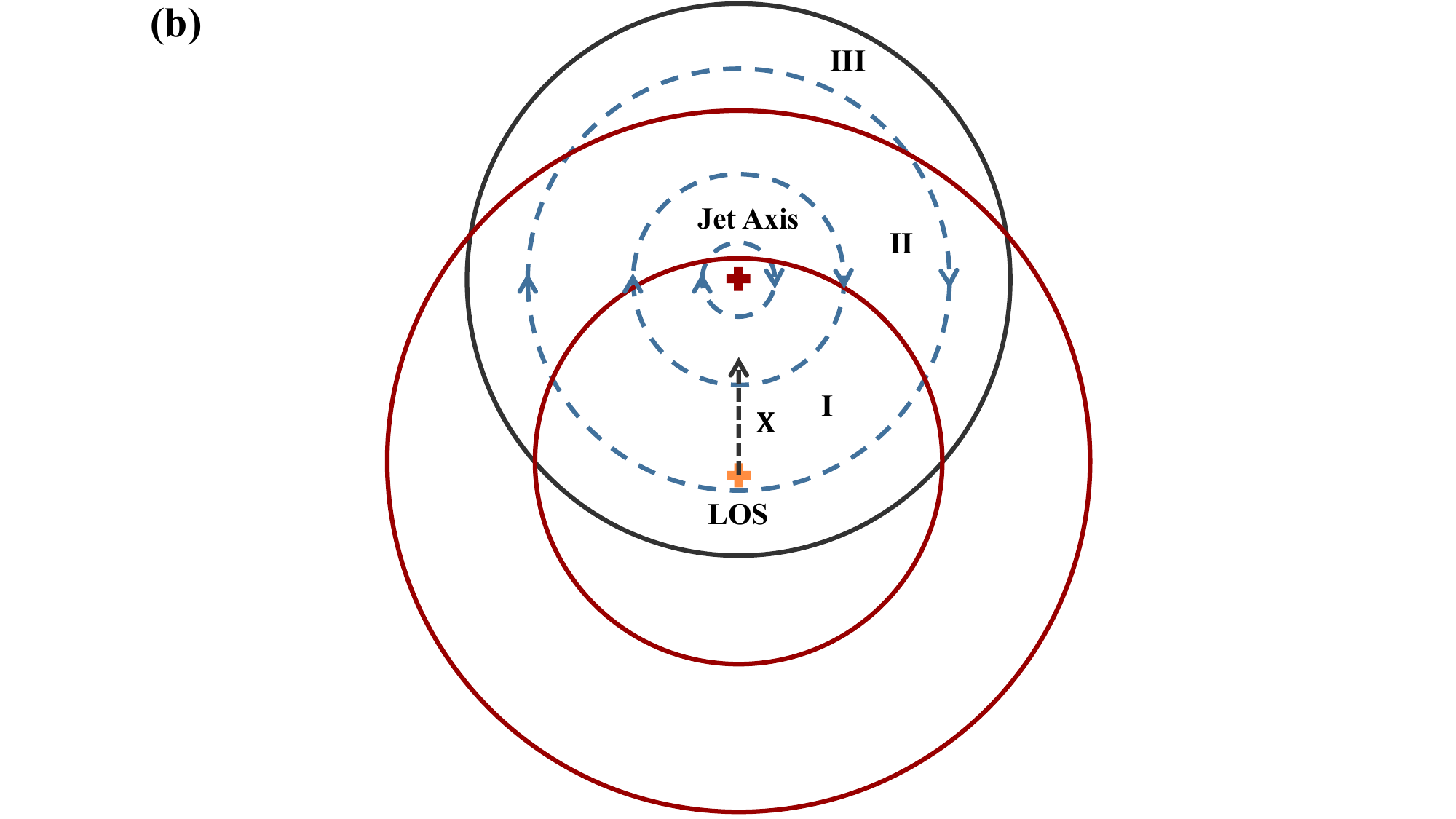}
	\end{minipage}
	\caption{Sketch of the observer observing the toroidal MFC at $\theta_\mathrm{V}$. The top panel (a) shows the geometry of the jet. The blue circles representing the toroidal MFC and direction. The bottom panel (b), the black dashed arrow represents the X-axis, I, II, and III are separated by two red circles. The PA of regions I and III is perpendicular to the X-axis, while the PA of region II is parallel to the X-axis.}
	\label{fig:Sketch}
\end{figure}

For giving a constant value of $\xi_\mathrm{B}$, the PD is highest during the burst time before $t_\mathrm{off}$, and it decreases after $t_\mathrm{off}$. Before the $t_\mathrm{off}$ point, the value of $\Pi$ exhibits a clear evolutionary trend with $\xi_\mathrm{B}$, as $\xi_\mathrm{B}$ increases, indicating a weaker random magnetic field strength, $\Pi$ also increases. Figure \ref{fig:variousepsilon} demonstrates that the evolution of the PD with respect to $\xi_\mathrm{B}$ in a given local point, assume the intrinsic PD be $\Pi_\mathrm{0}=0.8$ in the MFC under consideration, and the angle between the normal of that local point and the LOS is $0^\circ$, if the angle is not, when $\xi_\mathrm{B}$ is small enough，the random magnetic field will also generate polarization \citep{Laing1980}. When the value of $\xi_\mathrm{B}$ is less than 1, $\Pi$ exhibits a power-law growth with respect to $\xi_\mathrm{B}$, and the index is approximately 1.96, and the value of $\xi_\mathrm{B}$ is larger than 4, the PD approximates as $\Pi_0$, this means that when $\xi_\mathrm{B}$ is greater than a certain value, PD will not be affected by a random magnetic field. After the $t_\mathrm{off}$ point, apart from the instances the random magnetic field, we observe the PAs rotating twice by $90^\circ$, and the decay process remains consistent. This consistency is due to the intricate interplay between the curvature effect and the magnetic field structure of the photons emission region. Therefore, during this process, if the magnetic field structure remains the same, the temporal evolution of PDs at different $\xi_\mathrm{B}$ values after $t_\mathrm{off}$ remains unchanged. On the other hand, the changes in PAs are due to changes in the observation area (see Figure \ref{fig:Sketch}), the PA of regions I and III is perpendicular to the X-axis, while the PA of region II is parallel to the X-axis, before the first flip, the observed PA is dominated by the I region. A flip in PA means that the dominant region being observed has changed. And the timing of the PA's rotation by $90^\circ$ has been altered, as $\xi_\mathrm{B}$ decreases, the first flipping time occurs earlier, while the second flipping time has a relatively small impact, indicating that the time during which the PA is perpendicular to the $\hat{X}$-axis increases. The first flip is only noticeable when $\xi_\mathrm{B}$ is very small. This is because the smaller the PD before $t_\mathrm{off}$, the faster the depolarization will be, resulting in early flip. Before the second flip, the PD is small, resulting in minimal changes to the PD caused by $\xi_\mathrm{B}$. Consequently, the time changes to be small. When $\xi_\mathrm{B}=0$, we can only observe a minute PD for the first time when other parameters undergo a flip. This is attributed to the breakdown of symmetry \citep{Ghisellini1999, Granot2003andrandom}. Similarly, if the observation angle falls within the range $1-\cfrac{1}{\Gamma\theta_\mathrm{J}}<q<1$, we would observe a significant PD and two rotation of PAs by 90°.

By comparing different $\xi_\mathrm{B}$ lines, we can find that the more ordered the magnetic field is, the greater the PD becomes. When $\xi_\mathrm{B}>5$, the strength of the random magnetic field is weak, the results obtained using our proposed method consistently match those derived from calculations based on the formula for the purely ordered MFC, only slightly smaller than the ordered magnetic field before $t_\mathrm{off}$, e.g., $PD_\mathrm{\xi_\mathrm{B}=20}\approx PD_\mathrm{ord}$, so that when $\xi_\mathrm{B}=20$ the mixed magnetic field can be regarded as purely ordered magnetic field, and we have selected $\xi_\mathrm{B}=\xi_\mathrm{B, ord}=20$ as the purely ordered magnetic field.

\begin{figure}
    \centering
    \includegraphics[scale=0.35]{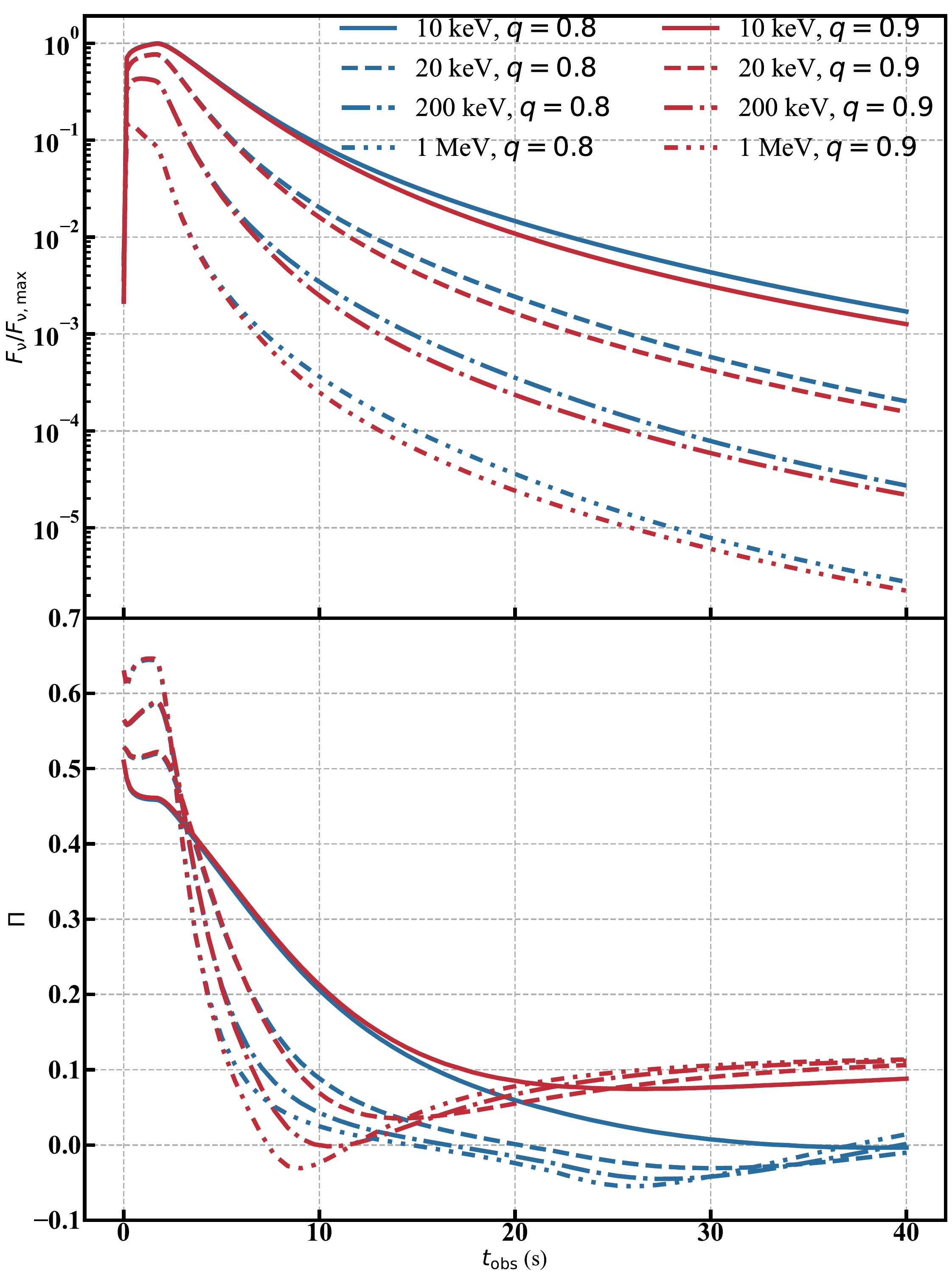}
    \caption{Time-evolved polarization in various observations of the energy. The top and bottom panels show the normalized light curves were obtained by dividing the original light curves by $F_\mathrm{\nu,max}$, which represents the maximum value of the 10 keV light curve, and the corresponding instantaneous PD curves, respectively. The MFC is toroidal magnetic field, and $\xi_\mathrm{B}=\xi_\mathrm{B,ord}$. The solid, dashed, dash-dotted, and dash-dot-dotted lines represent the scenarios of $h\nu_\mathrm{obs}$ at 10 keV, 20 keV, 200 keV, and 1 MeV, respectively. The blue and red lines correspond to the scenarios where $q=0.8$ and $q=0.9$, respectively.}
    \label{fig:ofT}
\end{figure}

In order to explore the relationship between PD and the different observations of the energy and the observing angles $q$. The observed energies $h\nu_\mathrm{obs}$ have been selected to correspond to 10 keV, 20 keV, 200 keV, and 1 MeV. Additionally, two observing angles, $q = 0.8$ and $q = 0.9$, and $\xi_\mathrm{B}$ is set equal to $\xi_\mathrm{B,ord}$, have been considered. Figure \ref{fig:ofT} has shown the normalized light curves (the top panel) were obtained by dividing the original light curves by $F_\mathrm{\nu,max}$, which represents the maximum value of the 10 keV light curve, and the corresponding instantaneous PD curves (the bottom panel). The PD is linked to the q, which exerts minimal influence on the light curves, but significantly impacts the subsequent PD evolution process and the rotation of PA. For 1 MeV, the PD evolution process begins differently at 5 seconds, while for low-energy the separation time is delayed, and $90^\circ$ flips occurred both 0.8 and 0.9. Nevertheless, the lower the energy level, the more difficult it is to achieve a flip. And all PD curves only maintain the highest PD before point $t_\mathrm{off}$, but after point $t_\mathrm{off}$, PD rapidly decreases and becomes the lowest during the entire time period, then rises again, if the time is long enough, we will only see one point at the edge of the jet, the PD is reached to the maximum PD. Based on the above analysis and Figure \ref{fig:Sketch}, for the toroidal MFC, the PD evolution patterns can be broadly classified into three distinct modes \citep{Gill2021,Cheng2024}. Under the red line parameter, PA undergoes two $90^\circ$ flips at 1 MeV, while there is no significant flip in other energy bands, the second transition mode performs relatively weakly. And when $q=0.8$, similar characteristics were exhibited. This is due to the fact that after $t_\mathrm{off}$, the injection of electrons stops, high-energy electrons radiation photons, and rapidly transforming into low-energy electrons. The low-energy electrons still exist abundantly and radiation photons. Consequently, the low-energy cutoff time is relatively delayed compared to high-energy, and the influence of the curvature effect on the PD transition mode is reduced, causing the PD at 10 keV to skip the rapid decay and PA flip processes in the middle and directly transition to the final evolution mode. Therefore, the evolution of PD curve is closely related to the energy range. In addition, the PD evolution of the energy band between 10 keV and 1 MeV is always sandwiched between 10 keV and 1 MeV. Therefore, this paper will only present the PDs evolution of 10 keV and 1 MeV.

\begin{figure}
    \centering
    \includegraphics[scale=0.35]{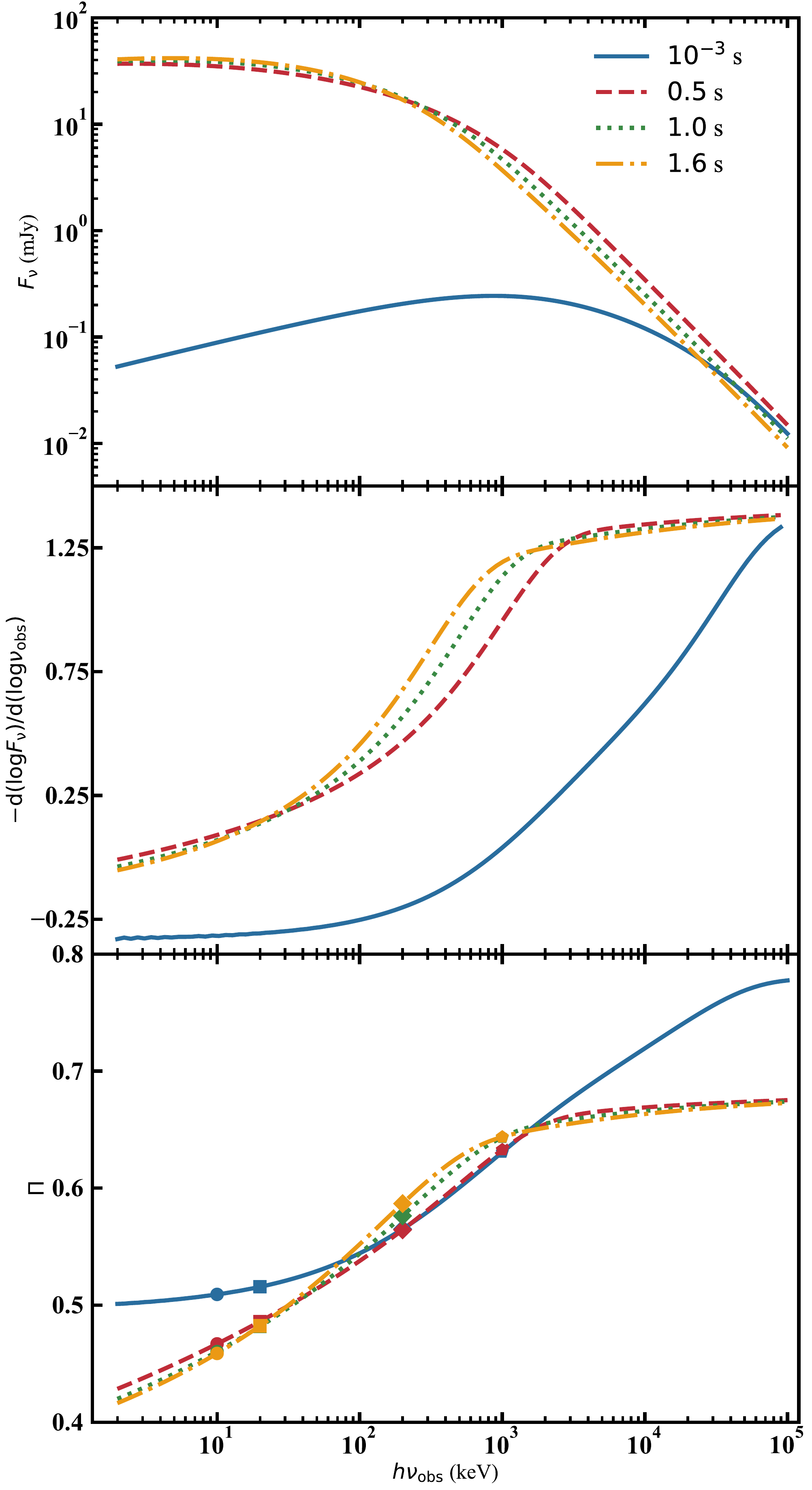}
    \caption{The polarization spectra were obtained under the condition of q=0.8. The top, middle, and bottom panels show the flux-density ($F_\mathrm{\nu}$) spectra, the negative spectral index of flux-density spectra, and PD spectra. The blue solid, red dashed, green dotted and yellow dash-dotted lines are for $10^{-3}$ s, 0.5 s, 1.0 s and 1.6 s, respectively. The observational energies of the circle, square, diamond and pentagon on the bottom panel are 10 keV, 20 keV, 200 keV and 1 MeV, respectively.}
    \label{fig:energyevolution}
\end{figure}

Both Figures \ref{fig:orderandrandom} and \ref{fig:ofT}, show that before $t_\mathrm{off}$ different energy bands have different evolutionary processes. The evolution of these PDs is closely related to the evolution of the energy spectrum. For the same parameters, we calculated the flux density ($F_\mathrm{\nu}$), the negative spectral index of flux density, and the PD spectra. The results are shown in Figure \ref{fig:energyevolution}. In this Figure, the observation times $t_\mathrm{obs}$ are set to $10^{-3}$ s, 0.5 s, 1.0 s, and 1.6 s. Specifically, the evolution of the initially injected power-law electron spectrum within a MFC, and magnetic field strength remains relatively constant due to its brief duration $t_\mathrm{obs}=10^{-3}$ s is discussed, along with the observation times 0.5 s, 1.0 s and 1.6 s prior to $t_\mathrm{off}$. Furthermore, the energies of 10 keV, 20 keV, 200 keV, and 1 MeV are marked with circles, squares, diamonds, and pentagons, respectively. 
For the initial PD spectra ($10^{-3}$ s), the PD value increases with increasing energy, and finally reaching a maximum of $PD_\mathrm{max} = \cfrac{p+2}{p+\frac{10}{3}} \approx 0.78$ at energies exceeding 100 MeV. Because when $\gamma'_\mathrm{e} $ exceeds $\gamma_\mathrm{min}'$, the electronic spectral index becomes $p+1$, and the minimum energy that reaches the $PD_\mathrm{max}$ is closely related to the magnetic field strength, $\gamma_\mathrm{min}'$, and the bulk Lorentz factor. On the other hand, a minimum value of $PD_\mathrm{min} \approx 0.5$ is observed at energies below 2 keV. This minimum value corresponds to the contribution of the electrons with the lowest energy, denoted as $\gamma_\mathrm{e,min}$, at that particular moment. The negative spectral index $m$ and the PD are related by $\Pi = \cfrac{m+1}{m+\frac{5}{3}}$, this relationship should also hold true for other time periods. However, the other time periods do not fully satisfy this condition, but rather align with the trend of the spectral index. This is because at $t_\mathrm{obs}=10^{-3}$ s, the observed magnetic field can be approximated as a point, whereas at later times, the observed magnetic field is treated as a surface, and it exhibits curvature. Consequently, geometric effects lead to depolarization, and the longer the time, the greater the depolarization becomes, leading to a decrease in $PD_\mathrm{max}$. And, over time, The accumulation of low-energy electrons and the decay of magnetic strength can lead to a decrease in high-energy radiation and an increase in low-energy occurrences in the energy spectrum, thereby causing energy spectrum shifts toward lower energy levels and a corresponding decrease in the minimum energy necessary to achieve $PD_\mathrm{max}$. But maintaining the changing trend of the spectral index at the same moment.

In this way, from the bottom panel of Figure \ref{fig:energyevolution}, we can understand the trend of changes in each energy segment before $t_\mathrm{off}$. For example, at 10 MeV, we expect PD to remain consistent after a rapid decline, and from 90 keV to 1 MeV, there will increase over time.

\begin{figure}
    \centering
    \includegraphics[scale=0.35]{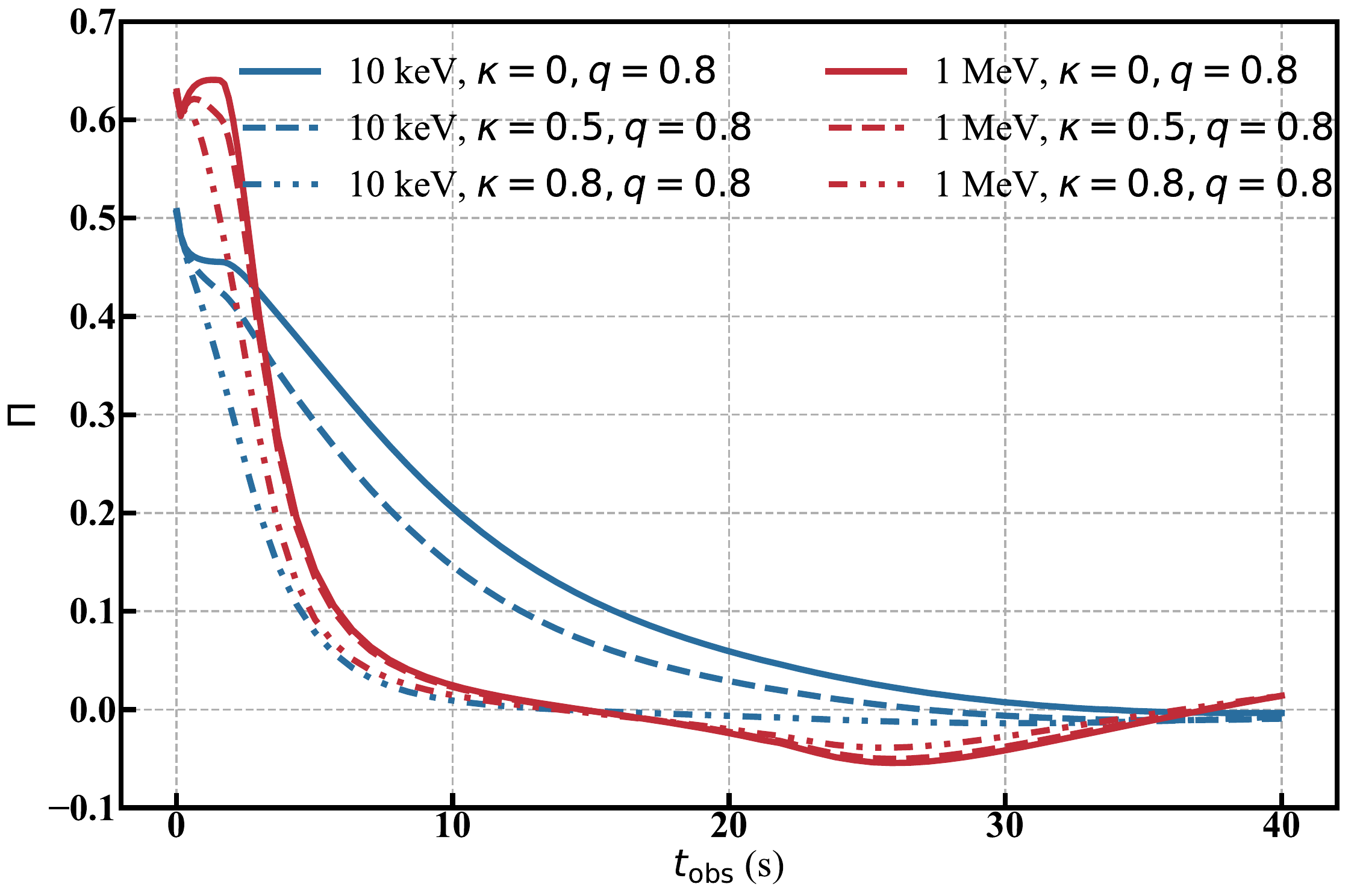}
    \caption{With the evolution of $\xi_\mathrm{B}$, observe the temporal evolution of PDs, while maintaining an observation angle of q=0.8, and the toroidal magnetic field. The blue and red lines correspond to observational energies of 10 keV and 1 MeV, respectively. The solid, dashed and dash–dot–dotted lines are for $\kappa=$ 0, 0.5 and 0.8, respectively.}
    \label{fig:variousB0.8}
\end{figure}

\begin{figure}
    \centering
    \includegraphics[scale=0.35]{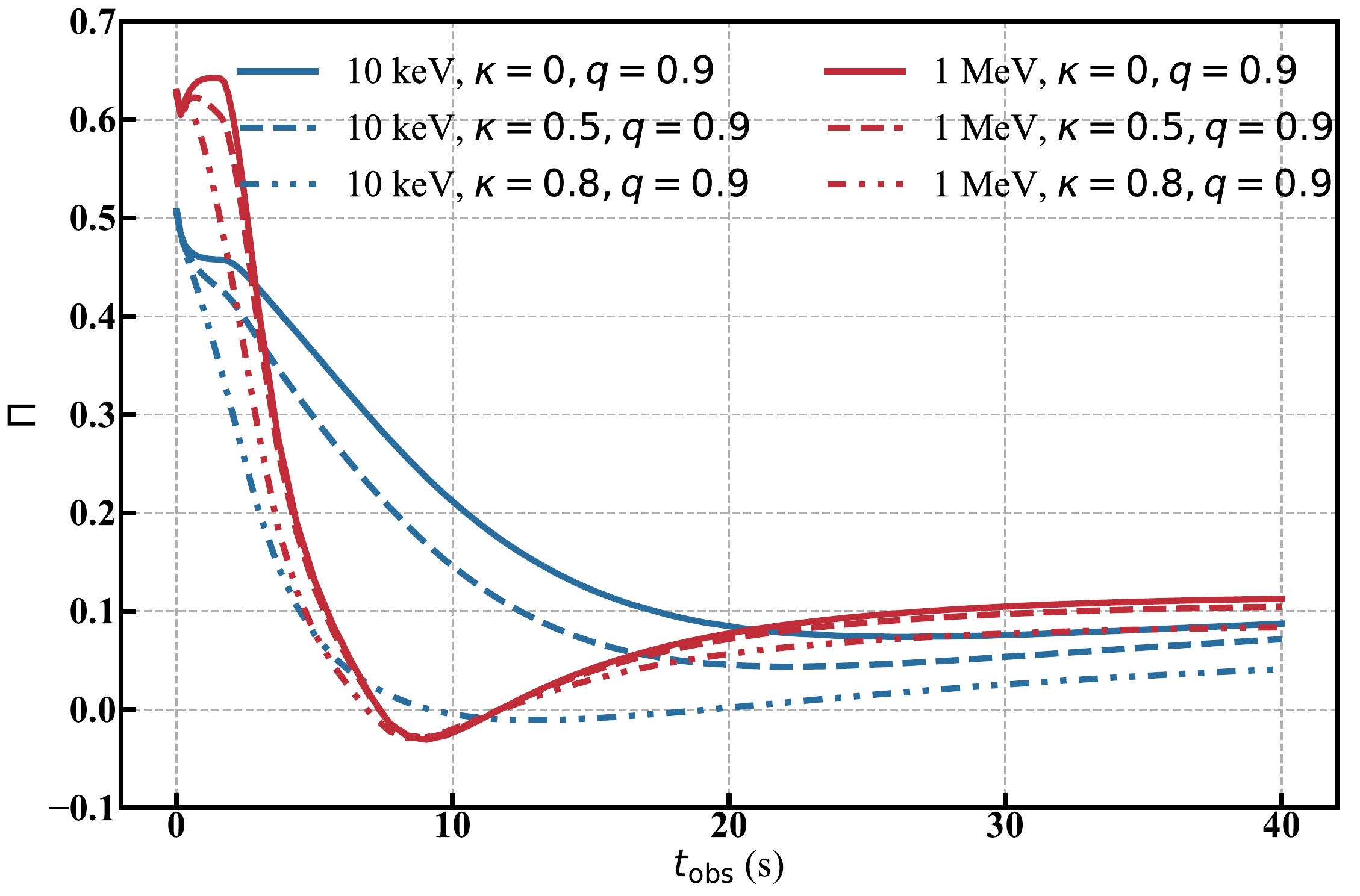}
    \caption{Same as Figure \ref{fig:variousB0.8}, but the observational angle $q=0.9$.}
    \label{fig:variousB0.9}
\end{figure}

For individual pulse in GRB, $\xi_\mathrm{B}$ decreases over time, i.e., the proportion of ordered magnetic field in the whole magnetic field decreases over time. The temporal evolution of PDs are shown in Figures \ref{fig:variousB0.8} and \ref{fig:variousB0.9}. The evolution formula for $\xi_\mathrm{B}$ is provided in equation \ref{eq:E}, with an initial value of $\xi_\mathrm{B,0}$ set to 10. The solid, dashed and dash–dot–dotted lines are for $\kappa=$ 0, 0.5 and 0.8, respectively. Additionally, the blue and red lines correspond to observational energies of 10 keV and 1 MeV, respectively. The change in MFC has little effect on the light curves, so the light curves are consistent with Figure 1, and the subsequent figures are also the same. In Figure \ref{fig:variousB0.8}, one can find that the impact of the value of $\kappa$ on PDs evolution is particularly pronounced for X-ray emission. The more rapidly $\xi_\mathrm{B}$ decays, the more drastic the changes in the PD curve become. In a well-ordered magnetic field, X-ray emissions do not exhibit PA flips. However, due to the introduction of a random magnetic field, low-frequency emissions also exhibit PA flips, when $\kappa=0.8$. For the high energy emission, the impact of the value of $\kappa$ on PD evolution is minimal, before $t_\mathrm{off} $ the PD decays with the decay of $\xi_\mathrm{B} $. And the high-energy emission ceases promptly due to a rapid decrease in high-energy electrons, resulting from the cessation of electrons injection at $t_\mathrm{off}$. The subsequent PD evolution is dominated by curvature effect, the $\xi_\mathrm{B}$ value of high-energy radiation seems to be frozen, so the hundreds of keV and MeV ranges maintain a relatively high $\xi_\mathrm{B}$, resulting in minimal variation. The reason for the significant changes in the X-ray range is the same, after the cessation of electrons injection at $t_\mathrm{off}$, the X-ray emission will not cease immediately, and during this period, the MFC continued to occur, only when significant radiation was stopped in the electronic spectrum at 10 keV, and the observed radiation was taken over by the curvature effect, with $\xi_\mathrm{B}$ not changing anymore. After $t_\mathrm{off}$, $\xi_\mathrm{B}$ has already diminished, thus the PD of the X-ray emission region exceeding $R_\mathrm{off}$ is approximately 0. The observing PD will primarily be influenced by regions exhibiting higher values of $\xi_\mathrm{B}$, which consistent with the hundreds of keV and MeV regime, and the observed polarization will tend to align with changes in the high-energy regime. In regions where the $\xi_\mathrm{B} $ is lower, especially for photons emitted after $t_\mathrm{off}$, their presence will only reduce the observed polarization. Therefore, when $\kappa=0.8$, the 10 keV and 1 MeV demonstrate such consistent trends of change, but the PD at 10 keV is lower. The lower $\xi_\mathrm{B}$ at $t_\mathrm{off}$, the smaller the difference between low energy and high energy. Figure \ref{fig:variousB0.9} is identical to Figure \ref{fig:variousB0.8}, except that the observational angle is $q=0.9$. We have also discovered that at 10 keV, $\kappa=0.8$, the evolution is similar to that observed in the high-energy range, and there are two occurrences of PA $90^\circ$ flips. 

\begin{figure}
    \centering
    \includegraphics[scale=0.35]{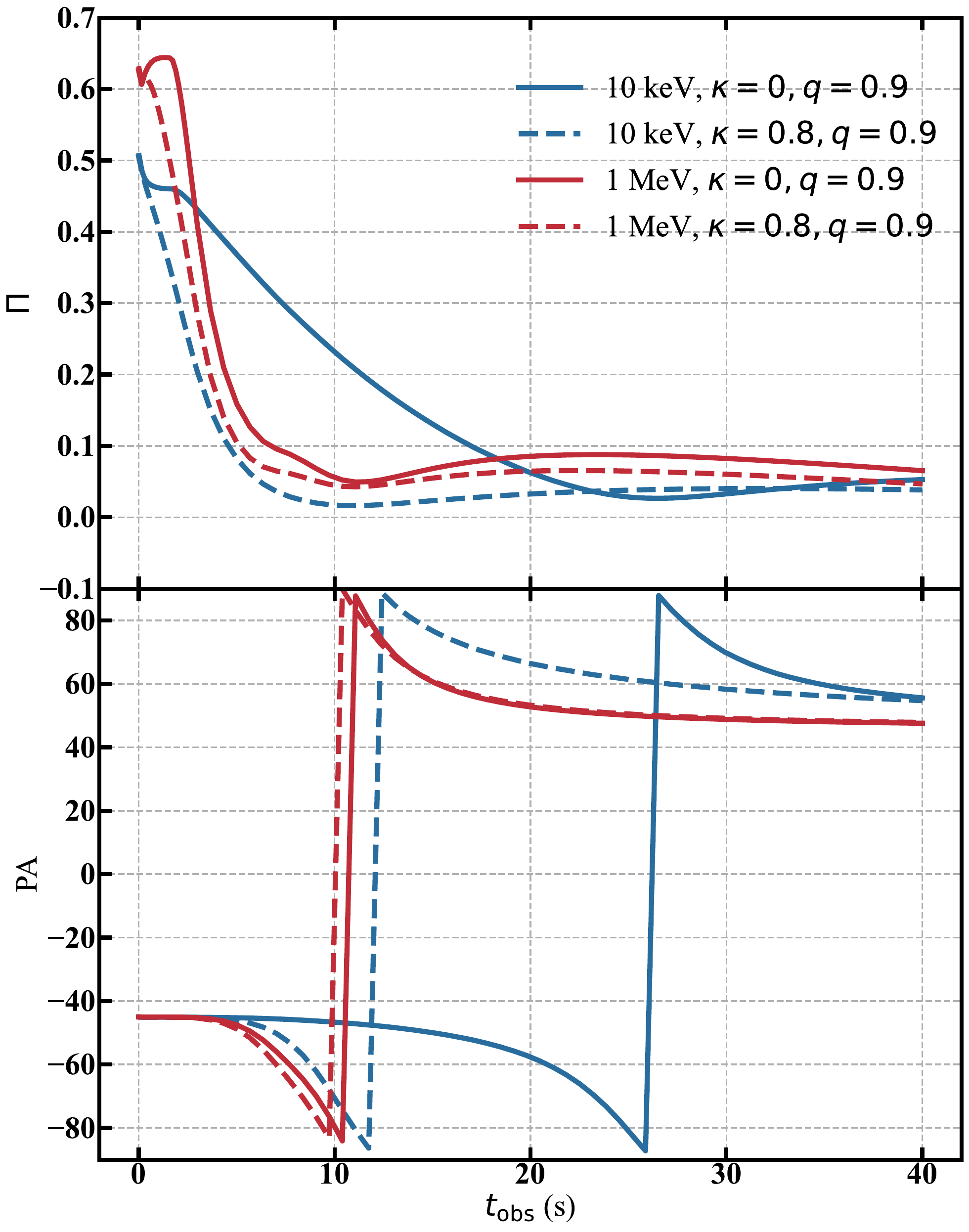}
    \caption{With the evolution of $\xi_\mathrm{B}$, observe the temporal evolution of PDs, while maintaining an observation angle of q=0.9, and the aligned magnetic field. The blue and red lines correspond to observational energies of 10 keV and 1 MeV, respectively. The solid and dashed lines are for $\kappa=$ 0 and 0.8, respectively.}
    \label{fig:A_0.9}
\end{figure}

Figure \ref{fig:A_0.9} show the aligned MFC. The behavior of random magnetic field on the evolution of PDs at any time is consistent with that of the toroidal MFC. A decrease in $\xi_\mathrm{B} $, i.e. a decrease in the ratio of ordered magnetic field strength to random magnetic field strength, will lead to a convergence of PDs in each energy band, making it difficult to distinguish. This is consistent with our analysis. 

\section{Discussion and Conclusions}\label{sec:Conclusions}
We considered a crude model, in which the magnetic field is initially ordered during the prompt emission within the shell. Following magnetic reconnection and turbulence, there is a transition to magnetic disorder, during which the MFC varies with time. The PD observation generated by the disordered magnetic field will be different from that generated by the pure ordered magnetic field in terms of the evolution process over time. Finally, these changes in MFC will affect the observed PD. 

The PD evolution behavior discussed in this paper is consistent with the general expectations of the ICMART model. The temporal evolution of PD depends on the details of the transition from ordered magnetic field to random magnetic field. Nevertheless, the results of the PD evolution are derived from Equation \ref{eq:E}. Consequently, the precise form requires additional numerical simulations for an accurate portrayal of the magnetic field fluctuations. However, this additional requirement does not affect the general conclusion.

Only when $\xi_\mathrm{B}$ is less than a certain value will PD significantly change. In the process of calculating the generated by a local region whose normal direction is consistent with the LOS, while other conditions remain unchanged, we observed that when $\xi_\mathrm{B}$ is less than 5, the PD value undergoes significant variations. Conversely, when $\xi_\mathrm{B}$ exceeds 10, it can be safely considered as an ordered magnetic field. When $\xi_\mathrm{B}<1$, the PD change could be approximated as a power-law function.  

If $\xi_\mathrm{B}$ gradually decays, the value of $\xi_\mathrm{B}$ at the moment when the electron stops injecting will determine whether the PD evolution at the low-energy X-ray light curve generated by the curvature effect is consistent with the high energy or greater than the polarization in the high energy range. Assuming that the evolution of $\xi_\mathrm{B}$ is described by equation \ref{eq:E}, we utilize this equation to estimate the observed PD. In this situation, we discovered that the presence of a random magnetic field significantly impacts the polarization across all energy ranges, with a greater effect observed in lower energy ranges. For a pure ordered magnetic field, after $t_\mathrm{off}$ the PD in the high-energy segment changes rapidly, while the change in the low-energy segment is relatively slow. Therefore, in the later stage of prompt emission, the PD in the low-energy segment will be higher than that in the high-energy segment. After considering the variation of $\xi_\mathrm{B}$, we observed that the polarization evolution behaviors are greatly related to the value of $\xi_\mathrm{B}$ at $t_\mathrm{off}$, the smaller the value, the more similar the trends shown by the low-energy and high-energy bands. When $\xi_\mathrm{B,0}=10$ and $\kappa=0.8$, the low-energy and high-energy bands showed similar trends, with only differences in PD values. And the decrease in $\xi_\mathrm{B}$ would cause the $90^\circ$ flip of PA to occur earlier, and it would also result in the emergence of PA changes in the PD curves at low energy ranges, which originally exhibited no PA variation under the original orderly conditions. This means that if a $90^\circ$ rotation of PA occurs in the high-energy range, the low-energy range may also exhibit similar behavior. In addition, these phenomena are more related to the participation of random magnetic fields, rather than the structure of ordered magnetic fields, whether they are toroidal or aligned.

We expect that the Low-Energy X-ray Polarization Detector (LPD, 2-10 keV) and the High Energy Polarization Detector (HPD, 30-800 keV) in POLAR-2 will observe similar evolution patterns in PD when $\kappa$ is large, i.e., when the ordering rapidly transitions to disorder, and the PD values will be slightly lower in this case. On the other hand, when $\kappa$ is small, we anticipate that the low-energy PD will be lower than the high-energy PD in the early stages, but the opposite will occur in the later stages. For the low-energy range, the PA rotation phenomenon, which was originally difficult to observe, will undergo significant changes due to the influence of a random magnetic field. The earlier this change occurs, the greater the photon count rate, thereby increasing the likelihood that LPD will detect this change. And we can utilize these detected results to ascertain the $\xi_\mathrm{B}$ following the termination of electron injection. 

\section*{Acknowledgements}
This work is supported by Department of Physics and GXUNAOC Center for Astrophysics and Space Sciences, Guangxi University. This work is supported by the National Natural Science Foundation of China (grant Nos. 12027803, U1731239, 12133003, 12175241, U1938201). This work is also supported by the Guangxi Talent Program (“Highland of Innovation Talents”).

\bibliography{sample631}{}
\bibliographystyle{aasjournal}

\end{CJK*}
\end{document}